\documentclass[11 pt]{amsart}
\usepackage{amsmath}
\usepackage{amssymb}
\usepackage{amsthm}
\usepackage{dsfont}
\usepackage{fancyhdr}
\usepackage[all]{xy}
\usepackage{hyperref}
\usepackage{paralist}
\usepackage{tikz}
\usepackage{verbatim}
\usepackage[hmargin=1.1in,vmargin=1.1in]{geometry}

\newtheorem{theorem}{Theorem}[section]
\newtheorem{lemma}[theorem]{Lemma}
\theoremstyle{proposition}

\theoremstyle{corollary}

\theoremstyle{definition}

\numberwithin{equation}{section}

\newcommand{\Z}{\mathbb Z}

\newcommand{\cL}{{\mathcal L}}
\newcommand{\F}{\mathbb F}

\newcommand{\D}{\mathbb D}
\newcommand{\Q}{\mathbb  Q}
\newcommand{\G}{\mathbb G}

\newcommand{\End}{\rm End}
\newcommand{\Pic}{\rm Pic}
\newcommand{\NS}{\rm NS}

\begin{document}
\title[]{Trilinear maps for cryptography}
\author[]{Ming-Deh A. Huang (USC, mdhuang@usc.edu)}
\address{Computer Science Department,University of Southern California, U.S.A.}
\email{mdhuang@usc.edu}

\urladdr{}

\maketitle

\begin{abstract}
We construct cryptographic trilinear maps that involve simple, non-ordinary abelian varieties over finite fields.  In addition to the discrete logarithm problems on the abelian varieties, the cryptographic strength of the trilinear maps is based on a discrete logarithm problem on the quotient of certain modules defined through the N\'{e}ron-Severi groups.  The discrete logarithm problem is reducible to constructing an explicit description of the algebra generated by two non-commuting endomorphisms, where the explicit description consists of a linear basis with the two endomorphisms expressed in the basis, and the multiplication table on the basis.  It is also reducible to constructing an effective $\Z$-basis for the endomorphism ring of a simple non-ordinary abelian variety. Both problems appear to be challenging in general and require further investigation.
\end{abstract}
\section{Introduction}
Cryptographic applications of multilinear maps were first proposed in the work of  Boneh and Silverberg \cite{BS}.  However the existence of cryptographically interesting $n$-linear maps for $n > 2$ remains an open problem. The problem has attracted much attention more recently as multilinear maps and their variations have become a useful tool for indistinguishability obfuscation. Very recently Lin and Tessaro \cite{LT} showed that trilinear maps are sufficient for the purpose of achieving indistinguishability obfuscation (see \cite{LT} for references to related works along several lines of investigation).

In this paper we study cryptographic trilinear maps involving abelian varieties over finite fields.
At the AIM workshop on cryptographic multilinear maps (2017) Chinburg suggested that the following map from \'{e}tale cohomology may serve as the basis of constructing a cryptographically interesting trilinear map:

\[ H^1 (A,\mu_{\ell})\times H^1 (A,\mu_{\ell})\times H^2 (A,\mu_{\ell})\to H^4 (A, \mu_{\ell}^\otimes{3})\cong \mu_{\ell}\]
where $A$ is an abelian surface over a finite field $\F$ and the prime $\ell\neq {\rm char}(\F)$.  This trilinear map is the starting point of our construction.

Suppose $A$ is a principally polarized abelian variety over a finite field $\F$. Let $\hat{A}$ denote the dual abelian variety. Consider $A$ as a variety over $\bar{\F}$, the algebraic closure of $\F$.  We have $H^1 (A,\mu_{\ell})\cong \hat{A}[\ell]\cong A[\ell]$.   We have $\Pic^0 A /\ell \Pic^0 A = 0$,  so $\NS(A)/\ell \NS(A)\cong \Pic (A)/\ell \Pic (A)$, where $\NS(A)=\Pic(A)/\Pic^0 (A)$ is the N\'{e}ron-Severi group .
From
\[ 0\to\mu_{\ell}\to\G_m\stackrel{\ell}{\to} \G_m\to 0\] and $H^1 (A,\G_m) = \Pic A$ we get
\[ 0\to  \Pic (A)/\ell \Pic (A) \to H^2 (A,\mu_{\ell}) \to H^2 (A, \G_m)[\ell] \to 0,\]
thus we can consider $\NS(A)/\ell\NS(A)$ as a subgroup of $H^2(A,\mu_{\ell})$
and we are led to a trilinear map
\[ A[\ell]\times A[\ell] \times \NS(A)/\ell \NS(A)\to \mu_{\ell}.\]
For an invertible sheaf $\cL$, let $\varphi_{\cL}$ be the map $A\to\hat{A}=\Pic^0 (A)$ so that
\[ \varphi_{\cL} (a) =t_a^*\cL \otimes \cL^{-1} \in \Pic^0 (A)\]
for $a\in A(\bar{\F})$ where $t_a$ is the translation map defined by by $a$ (\cite{MilneA} \S~1 and \S~6).
Let $e_{\ell}$ be the pairing between $A[\ell]$ and $\hat{A}[\ell]$ (\cite{MilneA} \S~16).
Then in the trilinear map, $(\alpha,\beta,\cL) \to e_{\ell}(\alpha,\varphi_{\cL} (\beta))$, where  $\alpha,\beta\in A[\ell]$, and $\cL$ is an invertible sheaf.

Note that in the map just described we no longer need to assume that $A$ is of dimension 2.

To construct a cryptographically interesting map, we need to work with the N\'{e}ron-Severi group more carefully.
We assume that $A$ is a simple, non-ordinary and  principally polarized abelian variety.

Suppose $\cL$ is an invertible sheaf associated to a Cartier divisor $D$.  Let $\varphi_D$ also denote $\varphi_{\cL}$.
Fix a divisor $\Theta$ such that $\varphi_{\Theta}$ is a principal polarization. Then
$\cL \to \lambda_D = \varphi_{\Theta}^{-1}\varphi_D$ determines an injection from $NS(A)$ to $\End (A)$, the endomorphism ring of $A$.   For $\alpha,\beta\in A[\ell]$, let $e_{\ell}^D (\alpha,\beta) = e_{\ell} (\alpha, \varphi_D (\beta))$.
Note that $e_{\ell}^D$ is skew-symmetric (\cite{MilneA} Lemma 16.2 (e)).  For any divisor $D'$ such that $\lambda_{D}=\lambda_{D'}$, we have $\varphi_D = \varphi_{D'}$, hence $e_{\ell}^D = e_{\ell}^{D'}$.

We choose a (random) divisor $D_1$ and find $\beta\in A[\ell]$ such that
$\varphi_{D_1} (\beta)=0$. For this we can choose a random $D$
such that the characteristic polynomial $f(x)$ of $\lambda_D = \varphi_{\Theta}^{-1}\varphi_{D}$ has a non-zero root $a$ mod $\ell$,  therefore
$f(x)=(x-a) f_1 (x) \mod \ell$, for some polynomial $f_1 (x)\in\F_{\ell} [x]$.   Let $\lambda=\lambda_D$. Replacing $f$ by a factor of $f$ if necessary we assume that $f(\lambda)=0\mod\ell$ but $f_1 (\lambda) \neq 0\mod\ell$.  Choose a random $\gamma\in A[\ell]$ so that $(f_1 (\lambda)) (\gamma) \neq 0$, and let $\beta=(f_1 (\lambda)) (\gamma)\in A[\ell]$.  Then
\[ (\lambda -a) (\beta) = ((\lambda - a) f_1(\lambda)) (\gamma) = (f(\lambda)) (\gamma)=0.\]
Observe that $\lambda-a = \varphi_{\Theta}^{-1}\varphi_{D-a\Theta}$.  So let $D_1 =D-a\Theta$.
Then $\varphi_{D_1} (\beta)=0$ as desired, and we have $e_{\ell}^{D_1} (\alpha,\beta) =1$ for all $\alpha\in A[\ell]$.

With $\beta$ and $D_1$ chosen, we choose another random $D_2$.
Let
$\lambda_{D_2} (\beta) = \alpha$.
 We have
\[ e_{\ell}^{D_2} (\alpha,\beta) = e_{\ell} (\alpha, \varphi_{D_2} (\beta))
= e_{\ell}(\alpha,\varphi_{\Theta} (\alpha)) = e_{\ell}^{\Theta} (\alpha,\alpha)=1.\]

In choosing $D_2$ we also make sure that $\lambda_{D_1} (\alpha)\neq 0$.  This implies
$\lambda_{D_1}\lambda_{D_2}\neq\lambda_{D_2}\lambda_{D_1}$
since $\lambda_{D_2}\lambda_{D_1} (\beta) = 0$ and $\lambda_{D_1}\lambda_{D_2} (\beta) = \lambda_{D_1}(\alpha) \neq 0$.
It also follows that $U/\ell U$ is of dimension 2.

Let $E$ be the submodule of $\End A$ containing all $\lambda_D$ where $D$ is a divisor.  Let $U$ be the $\Z$-submodule of $\End (A)$ generated by $\lambda_{D_1}$, $\lambda_{D_2}$, and the elements of $\ell E$.
Let $U_1$ be the $\Z$-submodule of $\End (A)$ generated by  1 and the elements of $U$.

If $D'$ is a divisor such that $\lambda_{D'}\in U$, then $\lambda_{D'}=\lambda_D$ for some divisor $D=x D_1 + y D_2 + \ell D_3 $ with $x,y\in\Z$ and $D_3$ a divisor.
Since
$e_{\ell}^{D_1} (\alpha,\beta) = e_{\ell}^{D_2} (\alpha,\beta) = e_{\ell}^{\ell D_3} (\alpha,\beta) =1$, we have
$ e_{\ell}^D (\alpha,\beta) =  1$ .
Since
$e_{\ell}^D=e_{\ell}^{D'}$, we have $e_{\ell}^{\Theta} (\alpha,\lambda_{D'}(\beta))=e_{\ell}^{D'} (\alpha,\beta)=1$.
So for $\lambda \in U_1$, $\lambda$ encodes 0 if and only if $\lambda \in U$ if and only if
$e_{\ell}^{\Theta} (\alpha, \lambda (\beta))=1$.

If $D'$ is a divisor such that $\lambda_{D'}\in a+U\subset U_1$ for some integer $a$, then $\lambda_{D'}=\lambda_D$ for some divisor $D=a\Theta + D_1$ with $\lambda_{D_1}\in U$.
Since $e_{\ell}^{D_1} (\alpha,\beta) =1$, we have $e_{\ell}^D (\alpha,\beta) = e_{\ell}^{a\Theta} (\alpha,\beta) =\zeta^a$ where
$\zeta = e_{\ell}^{\Theta} (\alpha,\beta)$.
Since
$e_{\ell}^D = e_{\ell}^{D'}$, we have $e_{\ell}^{D'} (\alpha,\beta)=\zeta^a$.

Let $G_1$ and $G_2$ be respectively the cyclic groups generated by $\alpha$ and $\beta$, and
 $G_3 = U_1/U$ with $1+U$ as the generator, we consider the trilinear map $G_1\times G_2 \times G_3 \to \mu_{\ell}$ sending
$(x\alpha,y\beta,z+U)$ to $\zeta^{xyz}$.

Following the cryptographic literature, we write
$[a]_i$ for an encoding of $a\in\Z/\ell\Z$ in $G_i$ for $i=1,2,3$.  For $a\in\Z/\ell\Z$, $[a]_1$ is the point $a\alpha$ and $[a]_2$ is the point $a\beta$.  In particular the encoding of $a\in\Z/\ell\Z$ in $G_i$ is deterministic for $i=1,2$.  In contrast the encoding in $G_3$ is probabilistic: for $a\in\Z/\ell\Z$, $[a]_3$ is $\lambda_D$, given in the form of a program $P_D$, where $D$ is a divisor
such that $\lambda_D \in a+U$.
The length of description of $P_D$ is polynomially bounded in the length of the description of $D$, and
$D$ is constructed to be linearly equivalent to $a\Theta + x D_1 + y D_2 +\ell D_3$ for some randomly chosen $x,y, D_3$ where $x,y\in\{0,\dots,\ell-1\}$ and $D_3$ is a divisor (see \S~\ref{rep} for details).

Given $[x]_1 = x\alpha$, $[y]_2 = y\beta$ and $[z]_3 = \lambda$ such that $\lambda\in z+U$, the trilinear map can be computed as $e_{\ell}^{\Theta} (x\alpha, \lambda(y\beta))=\zeta^{xyz}$ where $\zeta = e_{\ell}^{\Theta} (\alpha,\beta)$.

Suppose the Riemann-Roch space defined by a divisor is efficiently constructible, and the pairing $e_{\ell}^{\Theta}$ is efficiently computable.   We will show in \S~\ref{efficient} that under these assumptions the trilinear map is efficiently computable.

The cryptographic strength of the trilinear map is linked to the hardness of the discrete logarithm problem in the groups involved.
The one group that needs special attention is $G_3 = U_1/U$.
In the discrete logarithm problem for $G_3 = U_1/U$, given $\lambda \in U_1$ we are to determine $a$ such that $\lambda\in a+U$.

In the cryptographic setting we assume that polynomially many instantiations of $[1]_3$ are known.   From these encodings of 1 polynomially many divisors $\lambda \in U$ can be obtained.
Note that for $i=1,2,3$ and $a,b\in\Z/\ell\Z$, $a[b]_i$ is an encoding of $ab$ in $G_i$.

To investigate the hardness of the discrete logarithm problem on $G_3 = U_1/U$, it will be useful to consider the following  more general formulation of a discrete logarithm problem concerning the N\'{e}ron-Severi group $\NS(A)$.

 Fix a principal polarization $\varphi_{\Theta}$ determined by an ample divisor $\Theta$ with $\chi (\Theta)=1$.  Let $\iota: \NS(A) \to \End (A)$ be the injective map determined by $\varphi_{\Theta}$ under which the class of an invertible sheaf $\cL (D)$ associated to a Cartier divisor $D$ is mapped to $\lambda_D = \varphi_{\Theta}^{-1} \varphi_{D}$.  Let $E=\iota (\NS(A))$.

We use $||\:||$ to denote the bit-length in specifying a number or an object, whereas $|\:|$ denote the absolute value of a real number or the cardinality of a set.  Thus $|| A ||$ is the bit-length of the description of $A$, including the addition morphism $m_A$.  We assume that $m_A$ is {\em effectively specified} in the sense that given a point $a,b\in A(\bar{\F})$, $m_A (a,b)$ can be computed from the description of $m_A$ in time polynomially bounded in $||a||$ and $||b||$.  We assume that $|| A ||$ and $||\Theta||$ are polynomially bounded in $\log|\F|$ when $g=\dim A$ is fixed.

Suppose $M$ is a submodule of $E$ such that  $1\not\in M$.  Let $M_1$ be the submodule generated by 1 and the elements of $M$.  An element $\lambda\in M_1$ is presented as a program that on input a point $\alpha$ of $A$ computes $\lambda(\alpha)$ in time polynomially bounded in $||\lambda||$ and $||\alpha||$.  We assume that polynomially many $\lambda\in M$ can be randomly sampled where $||\lambda||$ is polynomially bounded in $||A||$ (hence in $\log |\F|$ when $g$ is fixed).

The discrete logarithm problem on $(M_1/M)\otimes \Z/\ell\Z$ is: given $\lambda \in M_1$, to determine $a\in\Z$ such that $\lambda - a =0$ in $(M_1/M)\otimes \Z/\ell\Z$.

In \S~\ref{DL} we consider various attacks on the discrete logarithm problem.

We show that if $M_1 = \Z \oplus M$, then the discrete logarithm problem can be effectively solved.  This is why in our construction the module $U$ contains $\ell E$, so at prime $\ell'\neq \ell$, $U/\ell' U = U_1/\ell' U_1 = E/ \ell' E$, consequently
$U_1 \neq \Z \oplus U$.

We show that, if
$M/\ell M$ is generated by mutually commuting elements, then the discrete logarithm problem can be effectively solved.   This is why we construct $\lambda_{D_1}$ and $\lambda_{D_2}$ in $U$ where $\lambda_{D_1}$ and $\lambda_{D_2}$ do not commute, and $U/\ell U$ is of dimension 2.

We show that the discrete logarithm problem can be effectively solved if
$M$ is contained in the center of the endomorphism algebra $\End^0 (A)=\End(A)\otimes\Q$.
The center of $\End^0 A$ is isomorphic to a CM field $\Q(\pi)$, where $\pi$ is a Weil number associated the Frobenius endomorphism.   We show that
the injective map of $E$ into $\Q(\pi)$ is efficiently computable, and with the injective map the discrete logarithm problem is reduced to straight-forward linear algebra.

The running times of these attacks are polynomial under reasonable heuristic assumptions, most notably that the bit-length of the characteristic polynomial for an endomorphism $\lambda$ is likely polynomially bounded in $||\lambda||$.

Our analysis shows that when $M/\ell M$ can be generated by commuting elements or when $M$ is contained in the center of $\End^0 A$, the discrete logarithm problem is tractable because we can work with a commutative subalgebra which can be  explicitly described.
Therefore we choose $A$ to be non-ordinary, so that $\End^0 (A)$ is a non-commutative division algebra. Moreover $E$, the image of $NS(A)$ in $\End A$, should not be contained in the center of $\End^0 (A)$.

In summary there are two important features about the group $G_3 = U_1/U$.
\begin{enumerate}
\item
Non-compatibility at $\ell'\neq \ell$: that $U_1 /\ell' U_1 = U/\ell' U = E/\ell' E$ whereas $U_1/\ell U_1 = \Z/\ell \Z \oplus U/\ell U$.
\item
Non-commutativity of algebra structure: $U_1$ is not contained in the center of $\End^0 A$ and $U/\ell U$ can  be generated by two elements in $\End A$ that do not commute.
\end{enumerate}

 Our construction leads to two interesting problems. If either problem can be solved efficiently for a simple non-ordinary abelian variety $A$ then trilinear maps constructed from $A$ are not secure.

The first problem is, given $\lambda,\mu\in E$ such that $\lambda\mu\neq \mu\lambda$, to construct an explicit description of the algebra $\Q[\lambda,\mu]$, by which we mean
a basis for $\Q[\lambda,\mu]$ as a vector space over $\Q$ with $\lambda$ and
$\mu$  expressed in the basis, and the multiplication table on the basis.

We say that an endomorphism $\lambda\in \End A$ is effectively specified if for $\alpha\in A(\bar{\F})$, $\lambda(\alpha)$ can be computed in time polynomial in $||\alpha||$ from the description of $\lambda$.

A basis $\mu_1$, ..., $\mu_m$ for a submodule $M$ of $\End A$ is {\em effective} if  $\mu_i$ is effectively specified for $i=1,\ldots, m$, moreover for every $\lambda=\sum_i a_i \mu_i \in M$ with $a_i\in\Z$, $|a_i |$ is polynomially bounded in $\deg\lambda$ for all $i$.

The second problem is to construct an effective $\Z$-basis for any submodule $M'$ of
$\End A$ containing $M_1$, in particular $M'$ can be $E$ or $\End A$.   It is an interesting question whether an effective basis for $E$ or $\End A$  exists and can be efficiently constructed.

These two problems have not been investigated in depth from an algorithmic perspective and appear to be quite challenging in general.

Fix as before a principal polarization $\varphi_{\Theta}$ and a corresponding injection $\iota$ from $\NS(A)$ to $\End A$.
The map $\iota$ naturally extends to an injection $\NS^0 (A) \to \End^0 A$ where $\NS^0 (A) = \NS(A)\otimes \Q$ and   $\End^0 A=\End A \otimes \Q$.  Through this map $\NS^0(A)$ is identified with the subspace $S$ of $\End^0 A$ whose elements are fixed by the Rosati involution defined by $\varphi_{\Theta}$.  As before let $\lambda_D = \varphi_{\Theta}^{-1} \varphi_D$ for divisors $D$.

Let $\D=\End^0 A$, a division algebra.  Let $K$ be the center of $\D$, and let $K_0$ be the subfield of $K$ consisting of elements fixed by the Rosati involution.
Let $d^2 =[ \D : K]= d^2$, $e=[K:\Q]$, $e_0 = [K_0 :\Q]$ and $\eta = \frac{\dim_{\Q} S}{\dim_{\Q} \D}$.
Abelian varieties can be classified into four types according to these numerical invariants (\cite{Mum} p.202). Non-ordinary abelian varieties are of Type II, III for IV, where $d \ge 2$.

If $A$ is not Type III, then $\eta \ge 1/2$, and $\dim_{\Q} S=\eta d^2 e \ge 2e \ge 2 e_0$.
In this case the image of a random element of $\NS(A)$ in $S$ is most likely not in the center $K_0$.
Consequently when choosing $D_1$ and $D_2$ to form $U$, it is very likely $\lambda_{D_1}$ and $\lambda_{D_2}$ are not in $K$.

When $A$ is Type III, $\eta = 1/4$, $d=2$, and $e=e_0$.  In this case $\dim_{\Q} S = e = e_0$, so $S\subset K$.
Therefore Type III abelian varieties are not adequate for our construction.

\section{What can be efficiently computed}
\label{efficient}
As before divisor will mean Cartier divisor, and since we are dealing with abelian varieties, we also think of them as Weil divisors.

For a divisor $D$, let $\cL (D)$ denote the invertible sheaf associated to $D$.  Let $k(V)$ denote the function field of a variety $V$.  Then

$H^0 (A, \cL (D)) = \{ f\in k(A) | (f) + D \ge 0\} \cup \{0\}$,
which we also denote as $L(D)$.

 As a Weil divisor, a divisor $D$ can be presented as a finite sum of prime divisors $\sum_{i=1}^n a_i v_i$ where $a_i\in\Z$ and $v_i$ is a prime of codimension 1.  The length of $D$, written $|| D ||$, is
$\sum_{i=1}^n (||a_i || + ||v_i||)$ where $||a||$ denotes the bit length of $a$ for $a\in\Z$, and $|| v_i ||$ is the length of the polynomials that define $v_i$.

For $\alpha\in A(\bar{\F})$, the bit length of $\alpha$, denoted $|| \alpha||$, is proportional to $n\log |\F'|$ where $\F'$ is the finite extension of $\F$ over which $\alpha$ is defined and $n$ the dimension of the ambient space in which the point is described.  We can consider $n$ a constant if the dimension of $A$ is fixed.

We assume that a basis of $L(D)$ can be computed in time polynomial in $|| D ||$ and the dimension of $L(D)$.  This is a reasonable assumption when dimension of $A$ is fixed.

\subsection{Computing $\lambda_D$}
We discuss how the map $\lambda_D = \varphi_{\Theta}^{-1}\varphi_D$ can be efficiently computed.  Assuming that the pairing $e_{\ell}^{\Theta}$ can be efficiently computed (which is the case for example when $A$ is the Jacobian of a curve), then
it will follow that the trilinear map described in the previous section can be computed efficiently.

\begin{lemma}
\label{unique}
Suppose $D$ is an effective divisor such that $\varphi_D$ is an isomorphism.  Then the only effective divisor linearly equivalent to $D$ is $D$ itself.
\end{lemma}
{\bf Proof}
Since $D$ is effective, $H^0 (A, \cL(D) )= L(D)\neq 0$, and since $\varphi_D$ is an isomorphism, it follows from Proposition 9.1 of \cite{MilneA} that $D$ is ample.   Since $\varphi_D$ is an isomorphism, $\deg\varphi_D =1$, and it follows from Theorem 13.3 \cite{MilneA} that $1=\chi (\cL(D)) = \dim H^0 (A,\cL (D))$.  Since $D$ is effective, this implies $L(D)$ contains only constant functions.  Therefore the only effective divisor linearly equivalent to $D$ is $D$ itself.  $\Box$

\begin{lemma}
\label{lambda}
Given a divisor $D$ and a point $a\in A(\bar{\F})$, $\lambda_D (a)$ can be computed in expected time polynomial in
$|| D ||$, $||\Theta||$ and $|| a ||$.
\end{lemma}
{\bf Proof}
For divisors $D$ and $D'$, $\varphi_{D + D'} = \varphi_{D} + \varphi_{D'}$.   Therefore
$\lambda_{D+D'} = \lambda_{D} + \lambda_{D'}$.  Hence if $D=\sum_i a_i v_i$ where $v_i$ is a prime divisor, then
$\lambda_{D} = \sum a_i \lambda_{v_i}$.

Therefore we may assume $D$ is a prime divisor.
 Observe that $\lambda_D (a) = b$ if and only if $\varphi_{\Theta} (b)=\varphi_D (a)$ if and only if $D_a - D \sim  \Theta_b - \Theta$ if and only if $D_a - D +\Theta\sim \Theta_b$.

Compute some $f\in L( D_a -D +\Theta)$, and let $D'=(f)+D_a-D+\Theta \ge 0$.  Then $D'\sim \Theta_b$, from Lemma~\ref{unique} we conclude that $D'=\Theta_b$.  The running time for computing $f$ is polynomial in $||D||$, $||\Theta||$ and $|| a ||$.  Note also that $D_a$, $D'$ and $b$ are defined over $\F'$ if $a\in A(\F')$.

From $\Theta_b$ we can determine $b$ as follows.  Sample a random finite set $S \subset \Theta$.  Then $b+\beta\in \Theta_b$ for all $\beta\in S$.   Solve for $x\in A$ such that $x+\beta \in \Theta_b$ for all $\beta\in S$.  This amounts to solving a polynomial system.  Note that $x+\beta\in\Theta_b$ implies $x\in\Theta_{b-\beta}$.  Hence
$x\in\cap_{\beta\in S} \Theta_{b-\beta}$.  When $S$ is large enough the intersection is likely of dimension zero, hence the polynomial system describing $x$ is likely of dimension zero, and can be solved efficiently when the number of variables is bounded.  One of the solutions for $x$ is $b$, and the correct $x$ can be tested by randomly choosing $\alpha\in\Theta$ and check if $\alpha+x \in\Theta_b$.  $\Box$

We remark that when $A$ is the Jacobian variety of a curve $C$, the problem can be solved even more efficiently, by reducing to constructing functions in the Riemann-Roch space of some divisor on $C$.

We assume that $\Theta$ has length polynomially bounded.  A divisor $D$ that is constructed in polynomial time also has length of description $|| D||$ polynomially bounded.    It follows from Lemma~\ref{lambda} that $\lambda_D$ is effectively specified.  Assuming $e_{\ell}^{\Theta}$ is efficiently computable, then the trilinear map $e_{\ell}^{\Theta} (u,\lambda_D (v))$ can  be computed in expected time polynomially bounded in $||u||$, $||v||$ and $||D||$.

\subsection{Computing the characteristic polynomial of an endomorphism}
Suppose $\lambda\in\End (A)$ is presented as a program that on input a point $\alpha$ of $A$ computes $\lambda(\alpha)$ in time polynomially bounded in $||\lambda||$ and $||\alpha||$.
Let $f\in\Z[x]$ be the characteristic polynomial of $\lambda$.  Then $f$ is of degree $2g$, where $g=\dim A$.
To determine $f$ it is sufficient by Chinese Remainder Theorem to determine $f\mod\ell'$ for sufficiently many small primes $\ell'$ with product greater than the maximum absolute value of the coefficients of $f$.  So after obtaining $f\mod\ell'$ for many $\ell'$, we have a candidate polynomial $f$ for the characteristic polynomial of $\lambda$.
We can check if $f(\lambda)=0$ by applying $f(\lambda)$ at a randomly chosen point and see if we get the zero point.

To determine $f\mod\ell'$ we first determine $\lambda\mod\ell'$ as a map on the $2g$-dimensional linear $\F_{\ell'}$-space $A[\ell']$, by constructing a basis $e_1,\ldots,e_{2g}$ for $A[\ell']$ and explicitly determining $\lambda (e_i)$ in terms of the basis.  This takes expected time polynomial in $(\ell')^{2g}$ and $||\lambda||$.    Then the characteristic polynomial of $\lambda \mod \ell'$ can be computed, hence $f\mod\ell'$.

Therefore we have the following

\begin{lemma}
\label{characteristic-polynomial}
\ \\
\begin{enumerate}
\item
For prime $\ell'$ not equal to the characteristic of $\F$, the map $\lambda \mod\ell'$ can be explicitly described in terms of a basis of $A[\ell']$ in expected time polynomial in $(\ell')^g$ and $|| \lambda ||$.
\item
The characteristic polynomial $f$ of $\lambda$ can be constructed in expected time polynomial in $|| \lambda ||$ and $|| f ||$.
\end{enumerate}
\end{lemma}

Recall that in forming $D_1$ we need to compute the characteristic polynomial of $\lambda_D$ for randomly chosen
divisor $D$.  By Lemma~\ref{lambda} $\lambda_D$ is effectively specified.  We make the heuristic assumption that for random $D$ it is likely that the characteristic polynomial $f_D$ has length polynomially bounded in $|| D||$, in which case $f_D$ can be constructed in expected time polynomial in $|| D||$.

\subsection{Constructing a random representative of a divisor class}
\label{rep}
Suppose $D$ is an effective divisor.  Since $A$ is simple, if $D$ is not ample then $\varphi_D = 0$, so
$\lambda_D =0$.  This can be tested by choosing a random $a\in A(\F)$ and check if $\lambda_D (a) =0$.

Now suppose $D$ is an effective ample divisor, and $m\in \Z_{> 0}$, we discuss how we can construct a random looking $D'\sim mD$ such that $|| D'||$ is polynomial in $|| m D ||=O(\log|m| ||D||)$.

Since $D$ is effective, $L(D)\neq \{0\}$.  Since $D$ is ample, by Theorem 13.3 \cite{MilneA} it follows that $\chi (\cL(D)) = \dim L(D) \ge 1$, and
$\dim L(sD) = \chi (\cL (sD)) = s^g \chi (\cL (D))\ge s^g$.   So the space of divisors linearly equivalent to $D$ has dimension $s^g -1 > 1$, when $s > 1$.

Choose small constant $s,t$ greater than 1, so that $m=sm_1 + t$ for $m_1 \in \Z_{\ge 0}$.
Choose a random $f_1 \in L(sD)$ and a random $f_2 \in L(tD)$.
Let $E_1 = (f) + sD$ and $E_2 = (f_2) + tD$.  Then
$mD\sim m_1 E_1 + E_2$ and $|| m_1 E_1 + E_2||$ is polynomially bounded in $|| mD ||$.

In our construction if a divisor $D$ is chosen in the form
$D=a\Theta + b D_1 + cD_2 + \ell D'$, we construct a random looking divisor $D''$ linearly equivalent to $D$ to encode the class of $\lambda_D + U = a+U$.

we can apply the above procedure to $a\Theta$, $bD_1$, $cD_2$ and $\ell D'$ respectively to construct $D'_0\sim a\Theta$, $D'_1\sim bD_1$, $D'_2\sim cD_2$ and $D'_3\sim \ell D'$.
Then $D''=D'_0+D'_1 + D'_2 + D'_3$ is linearly equivalent to $D$, so $\lambda_{D'}=\lambda_D$.

If $D$ is given in the form $D=\sum_i a_i v_i$ where $v_i$ is a prime divisor, we may apply the above procedure to each $v_i$ that is ample to construct some $D'_i\sim a_i v_i$. Then $\sum_i D'_i$ is linear equivalent to $D$.  The same process can be applied to $D'_i$ again, and by repeating this process sufficiently many times, we can construct a divisor $D'' = \sum_i b_i v_i \mod\ell$ where the sum involves many, though polynomially bounded in number, prime divisors $v_i$.
We have $\lambda_D = \lambda_{D''} = \sum_i b_i \lambda_{v_i}\mod\ell$.  Each $\lambda_{v_i}$ has a program $p_i$ of length polynomially bounded in $||v_i||$ by Lemma~\ref{lambda}.  The program for $\lambda_{D''}$ can be specified in length $\sum_i || b_i || +||p_i||$ , which is polynomially bounded in $|| D''||$, hence in $|| D ||$.

We have seen  by virtue of Lemma~\ref{lambda} that the map $\NS(A)/\ell NS(A) \to E/\ell E$ is efficiently computable. An interesting question is whether the inverse is efficient to compute as well.  That is, given $\mu\in E$, can we construct efficiently a divisor $D$ such that $\lambda_D = \mu\mod\ell$?   In light of the discussion in the next subsection, an affirmative answer would reduce the discrete logarithm problem that concerns us to intersection product.
The answer is in the affirmative when $\End A$ is commutative, and this will follow from Lemma~\ref{iota}.   However the situation in the non-commutative case is far from being clear.

\subsection{Linear algebra on $\NS(A)$ reduces to intersection product}
The reason why the discrete logarithm problem involving $\NS(A)$ is specified in terms of elements in $E$ is because linear algebra in $\NS(A)$ and $\NS(A)/\ell\NS(A)$ can be reduced to intersection product of divisors.    More specifically to determine a linear relation in $\NS(A)/\ell\NS(A)$ between a divisor $D$ and a finite set of divisors $D_1$, ..., $D_n$, we want to solve for $x_i$ such that   $D$ is algebraically equivalent to $\sum_i x_i D_i$ modulo $\ell$.  Let $g=\dim A$.  Observe that for a divisor $H$, the algebraic equivalence implies
$D\cdot H^{g-1} = \sum_i x_i D_i\cdot H^{g-1}\mod\ell$.  Therefore by computing the $(n+1)$ intersection products $a=D\cdot H^{g-1}$ and $b_i = D_i\cdot H^{g-1}$, for $i=1,\ldots,n$, we get a linear relation
$a=\sum_i b_i x_i \mod\ell$.  With sufficiently many linear relations we can determine $x_i$.  Therefore if the discrete logarithm problem is specified in terms of divisors then the problem can be reduced to computing intersection products.
If $g=\dim A$ is fixed, then intersection products on $A$ can in principle be reduced to counting solutions of polynomial systems in bounded number of variables.

\section{The discrete logarithm problem involving $NS(A)$}
\label{DL}
We discuss various attacks on the underlying discrete logarithm problem concerning the N\'{e}ron-Severi group.  Let us recall the general set-up of this problem.

Let $A$ be a principally polarized abelian variety defined over a finite field $\F_q$. Fix a principal polarization $\varphi_{\Theta}$ determined by an ample divisor $\Theta$ with $\chi (\Theta)=1$, and consider the injection $\iota$ of $\NS(A)$ to $\End (A)$ determined by $\varphi_{\Theta}$, such that the class of an invertible sheaf $\cL (D)$, where $D$ is a divisor, is mapped to $\lambda_D = \varphi_{\Theta}^{-1} \varphi_{D}$.

Suppose $M$ is a submodule of $E$ such that  $1\not\in M$.  Let $M_1$ be the submodule generated by 1 and the elements of $M$.  We do not assume that $M$ is explicitly given, however polynomially many elements of $M$ can be randomly sampled.

The discrete logarithm problem on $(M_1/M)\otimes \Z/\ell\Z$ is: given $\lambda \in M_1$, to determine $a\in\Z$ such that $\lambda - a =0$ in $(M_1/M)\otimes \Z/\ell\Z$.

\subsection{The case $M_1 = \Z \oplus M$}
Given $\lambda\in M_1$, we want to determine $a\mod\ell$ such that $\lambda = a + \mu$ with $a\in\Z$ and $\mu\in M$.  To determine $a$ it is sufficient to determine $a\mod\ell'$ for sufficiently many small primes $\ell'$.

We may assume that we have sampled enough elements of $M$ that they generate $M/\ell' M$ as a vector space over $\F_{\ell'}$.  By Lemma~\ref{characteristic-polynomial} we can determine for each sampled element $\mu$  the action of $\mu$ on a basis of $A[\ell']$ in time polynomial in ${\ell'}^g$.
Therefore we can construct a $\F_{\ell'}$-basis $\mu_1$, ..., $\mu_k$ of $M/\ell' M$ in time polynomial in ${\ell'}^g$.

For all $a_0$,..., $a_k\in\F_{\ell'}$, we can check if $\lambda = a_0 + a_1 \mu_1 + ..., a_k \mu_k\mod\ell'$ by
acting on $A[\ell']$ or a basis of $A[\ell']$.  Once the equality is verified we know that $a=a_0 \mod\ell'$. The amount of time required is polynomial in ${\ell'}^g$.

In constructing our trilinear map we make sure that for $\ell'\neq \ell$, $U/\ell' U= U_1/\ell' U_1$, in other words, $U$ and $U_1$ are indistinguishable mod $\ell'$.  This is accomplished by including $\ell E$ in  $U$.

\subsection{The case $M/\ell M$ is of dimension no greater than one}

Since $\deg$ is a (homogeneous) polynomial function of degree $2g$ on $\End^0 A$ (\cite{MilneA} Proposition 12.4), it follows that
$\deg (a\lambda) = a^{2g} \deg\lambda$ for $\lambda\in\End^0 A$, moreover
if $\lambda,\mu\in\End A$ and $\lambda=\mu\mod\ell$, then $\deg\lambda = \deg\mu\mod\ell$.
Hence the discrete-log problem in the case $M=\ell E$ is reduced to degree computation.  In particular if $\lambda=a \mod\ell$ then $\deg\lambda = a^{2g} \mod\ell$.  Determine the characteristic polynomials of $\lambda$.  From the constant terms of the polynomials we get $\deg \lambda$.  Then $a$ can be determined.

\subsubsection{$M/\ell M$ is of dimension 1}  In the discrete logarithm problem we have polynomially many samples of elements in $M$.  Pick one $\lambda$ such that $\lambda\neq 0 \mod\ell$, which can be checked by choosing a random $\alpha\in A[\ell]$ and verify that $\lambda (\alpha)\neq 0$.

Given $\mu \in M_1$ we want to find $a$ such that $\mu = a+ b\lambda \mod\ell$ where $b\in\Z$.

Since $a+b\lambda \in \Q [\lambda]\subset \End^0 A$ and $\Q [\lambda]$ is a commutative field, every root of the characteristic polynomial of $a+b\lambda$ is of the form $a+b \gamma$ where $\gamma$ is a root of the characteristic polynomial of $\lambda$.  The characteristic polynomial of $a+b\lambda$ is $f(\frac{x-a}{b})$ where $f$ is the characteristic polynomial of $\lambda$.

Compute the characteristic polynomial $g$ of $\mu$.  Then $g(x)= f(b^{-1} (x-a))\mod\ell$, where $g$ and $f$ are known and $a, b$ are unknown.
Comparison of each coefficient gives rise to a polynomial equation in $a$ and $b$ over $\F_{\ell}$.
Solving the system of polynomial equations we can determine $a$ and $b$ up to a finite number of choices.  Then determine the correct one by acting on a random point of $A[\ell]$.

\subsection{The case $M/\ell M$ is generated by at least two elements}
The line of attack described below can be easily generalized, however
for simplicity, we illustrate the ideas with the case where $M/\ell M$ is generated by two elements.  So from the random samples of elements of $M$ pick two, $\lambda$ and $\mu$, then it is likely that they generate $M/\ell M$.  Therefore the discrete logarithm problem can be described as follows: given $\omega \in M_1$, to find $a$ such that $\omega = a+ b\lambda+c\mu\mod\ell$ where $b,c\in\Z$.

By an explicit description of the algebra $K=\Q[\lambda,\mu]$ we mean a basis over $\Q$ as vector space with $\lambda$ and $\mu$ expressed in terms of the basis, and the multiplication table on the basis elements.

If $\lambda$ and $\mu$ commute,  then $K=\Q[\lambda,\mu]$ is a commutative field of extension degree dividing $2g$.  Let $f(x)$ and $g(x)$ be respectively the irreducible polynomials of $\lambda$ and $\mu$.
An explicit description of the algebra $K$ can be obtained from $f(x)$ and $g(x)$ as follows.
Factor $g(x)$ over $\Q[\lambda]$.  Find the irreducible factor $h(x)\in\Q[\lambda](x]$ of $g(x)$ such that $h(\mu)=0$, by choosing random points $w$ on the variety $A$ and checking if $h(\mu) (w)=0$.  The field $\Q[\lambda,\mu]$ is of extension degree $\deg f \deg h$ over $\Q$, with $\lambda^i\mu^j$, $0\le \deg f-1$, $0\le\deg h-1$ as a basis of $\Q[\lambda,\mu]$ over $\Q$.  The multiplication table on this basis can be written using $f$ and $h$.

Then we can express the action of $a+b\lambda+c\mu$ on the basis with the help of the multiplication table and from that determine the characteristic polynomial of $a+b\lambda+c\mu$ acting on $K$, with coefficients being polynomials in $a,b,c$.
This polynomial, if the degree is $2g$, or a suitable power of this polynomial, is the characteristic polynomial of $a+b\lambda+c\mu$ as an element of $\End^0 A$.  Then from $\omega = a+b\lambda+c\mu\mod\ell$ and by comparing the coefficients of the characteristic polynomials of $a+b\lambda+c\mu$ and $\omega$, we obtain a system of polynomials in $a,b,c$, from that we can determine $a$ as before.

However, if $\mu\lambda \neq \lambda\mu$, we run into difficulty
if we try to mount the same line of attack.  In this case $\Q [\lambda,\mu]$ as a subalgebra of $\End^0 A$ is not commutative and it is not clear whether one can efficiently determine the structure of $\Q[\lambda,\mu]$ explicitly.
This is why in our trilinear map the $\F_{\ell}$-dimension of $U/\ell U$ is generated by two elements that do not commute.

Below we give a reduction from the discrete logarithm problem to further clarify how the security of our trilinear map depends on the hardness of constructing an explicit description of the subalgebra generated by two non-commuting elements.

From the random samples of elements in $M$ pick two, $\lambda$ and $\mu$, then it is likely that they generate $M/\ell M$.  Therefore the discrete logarithm problem can be described as follows: given $\omega \in M_1$, to find $a$ such that $\omega = a+ b\lambda+c\mu\mod\ell$ where $b,c\in\Z$.

Suppose we are given a $\Q$-basis of $\Q[\lambda,\mu]$ with $\lambda$ and $\mu$ expressed in the basis, as well as the multiplication table on the basis.   We want to determine $a,b,c$ such that $\omega=a+b\lambda+c\mu\mod\ell$.  As before we can express the action of $a+b\lambda+c\mu$ on the basis with the help of the multiplication table and from that determine the characteristic polynomial $F$ of $a+b\lambda+c\mu$ acting on $K$, with coefficients being polynomials in $a,b,c$.  Let $\rho$ denote the irreducible polynomial of $a+b\lambda+c\mu$.
Let $f$ denote the characteristic polynomial of $a+b\lambda+c\mu$ as an element of $\End^0 A$.

Let $d_h$ denote the degree of a polynomial $h$.  Then $d_f=2g$, $d_{\rho} | 2g$ and $d_F \le [\End^0 A : \Q] \le 4g^2$.
We have $F = \rho H$, $f=\rho h$ for some polynomials $H$ and $h$. For each choice of $d_{\rho} | 2g$, we set up a polynomial system as follows. Treat $\rho$, $h$ and $H$ as unknown polynomials.  We have $d_F -d_{\rho} +2g +3$ unknown including $a,b,c$ and the unknown coefficients of $\rho$, $h$ and $H$.

Let $F_{\omega}$ be the characteristic polynomial of $\omega$, which is of degree $2g$.
From $F_{\omega} = f =\rho h\mod\ell$ and $F = \rho H$, by comparing coefficients, we derive $2g+d_F$ many polynomial equations.
If $d_{\rho} \ge 3$ then $2g+d_F \ge d_F -d_{\rho} +2g+3$. There are at least as many polynomial equations as the number of unknown variables, so we expect on heuristic ground to have finitely many solutions.   For each solution we check if $\omega = a+b\lambda+c\mu\mod\ell$ by acting on a random point in $A[\ell]$.

\subsection{The case $M$ is contained in the center of $\End^0 A$}  Let $C(A)$ denote the center of $\End^0 A$.  Then there is an isomorphism $\iota : C(A) \to \Q(\pi)$ where $\Q(\pi)$ is a CM field and $\pi$ is the image of the Frobenius endomorphism $\pi_A$ under $\iota$.

\begin{lemma}\label{iota}The restriction of the map $\iota$ to $E\cap C(A)$ is efficiently computable; that is, given $\lambda\in C(A)$, $\iota(\lambda)$ can be computed in time polynomial in $||\lambda||$ and $||f||$ and $||\pi||$, where $f$ is the characteristic polynomial of $\lambda$.\end{lemma}
\ \\{\bf Proof}  Suppose $\lambda \in C(A)$.  Then $\iota(\lambda)$ is a root of the characteristic polynomial $f$ of $\lambda$.  By Lemma~\ref{characteristic-polynomial} $f$ can be constructed in time polynomial in $|| \lambda ||$ and $||f ||$.  By factoring $f$ over $\Q(\pi)$, we can express each root of $f$ in $\Q(\pi)$ in the form
$m^{-1}\sum_i a_i \pi^i$ where $m$ and $a_i$ are integers. This takes time polynomial in $|| f ||$ and $|| \pi ||$.   We want to check if  $\lambda = m^{-1}\sum_i a_i \pi_A^i$.
If the equality does not hold then $m\lambda \neq \sum_i a_i \pi_A^i \mod\ell'$ with high probability for random prime
$\ell'$ not dividing $m\deg\lambda$.  Therefore
choose a small random $\ell'$ not dividing  $\deg\lambda$ and $m$, and check that the equality holds mod $\ell'$  by acting on $A[\ell']$, as in Lemma~\ref{characteristic-polynomial}.  Then the correct root of the characteristic polynomial of $\lambda$ can be determined as the image under $\iota$. $\Box$

Given polynomially many elements $\lambda$ in $M$, we can by Lemma~\ref{iota} efficiently compute their images under $\iota$ and
determine $\iota (M)$ as a subspace of $\Q (\pi)$.

Given $\lambda \in M_1$, we want to determine $a\in\Z$ such that $\lambda -a \mod \ell \in M/\ell M$.  This reduces to finding $a$ such that $\iota(\lambda)-a \mod\ell \in \iota(M)$, which can be solved efficiently since
linear algebra in $\Q(\pi)$ is easy once elements are expressed in terms of the basis $\pi^i$, $i=0$, ..., $e-2$ where $e=[\Q(\pi) : \Q]$.

If $A$ is simple and ordinary then $C(A)=\End^0 (A)$.  This is why we do not consider ordinary abelian varieties.

\subsection{The case when $M_1\not\subset C(A)$}
In this case the discrete logarithm appears to be difficult generally speaking.  However suppose we are given
a submodule $M'$ of $\End^0 A$ containing $M_1$ together with an effective
basis $\mu_1$, ..., $\mu_n$ of $M'$ as a $\Z$-module.
Then the discrete logarithm problem on $M_1/M \otimes \Z/\ell\Z$ can be solved efficiently, as we show below.

Since the basis for $M'$ is effective, for each $\mu_i$ in the basis and prime $\ell'$, the action of $\mu_i$ on $A[\ell']$ can be explicitly determined in time $(\ell')^{O(g)}$.   Moreover for every $\lambda=\sum_i a_i \lambda_i \in M'$ with $a_i\in\Z$, $|a_i |$ is polynomially bounded in $\deg\lambda$ for all $i$.

Let $\lambda\in M_1$.  We can find $a_i\in\Z$ such that $\lambda = \sum_{i=1}^n a_i \mu_i$ as follows.
For a small prime $\ell'$, find $b_i < \ell'$ such that $\lambda$ and $\sum_i b_i \mu_i$ act the same on $A[\ell']$.
By Lemma~\ref{lambda} and the assumption on $M'$ we know that this can be done in $(\ell')^{O(g)}$ time.

Then $a_i = b_i \mod\ell'$ for all $i$.  By the assumption on the basis of $M'$, we only need to consider all primes $\ell'$ up to a bound polynomial in $\log(\deg\lambda)$ in order to determine $a_i$ for all $i$ by Chinese Remaindering.

We have polynomially many $\lambda_i\in M$ and a $\lambda \in M_1$, and
we would like to find $a\in\Z$ such that $\lambda - a \mod\ell \in M_1/M \otimes \Z/\ell\Z$.

We can determine the decomposition of all $\lambda_i$, for all $i$, $\lambda$, and $1$ (as an endomorphism) in terms of the basis for $M'$ as above, and suppose
$u_i$, and $v$ and $v_0$ are respectively the coefficient vectors in their decomposition.
Suppose $u_i\mod\ell$ generate $M/\ell M$ as a vector space over $\F_{\ell}$, which is most likely the case.
Then $v=x_0 v_0 + \sum_{i=1}^n x_i u_i\mod\ell$ for some $ x_i \in \Z/\ell\Z$ for $i=0,\ldots,n$.  Solve the linear system over $\F_{\ell}$ to obtain a solution $x_i$, $i=0,\ldots,n$, and $x_0$ is what we want, that is
$a=x_0\mod\ell$.

\subsection{Summary}
In constructing the trilinear map we choose $A$ to be a simple but non-ordinary abelian variety with principal polarization.  As discussed above, the construction gives rise to a discrete logarithm problem on some $U_1/U \otimes \Z/\ell\Z$ where $U/\ell U$ has a basis of two elements that do not commute, and $U_1/\ell' U_1 = U/\ell' U$ for primes $\ell'\neq \ell$.

The discrete logarithm problem is reducible to constructing an explicit description of the algebra generated by two non-commuting endomorphisms - a basis over $\Q$, the multiplication table for the basis, and the two elements expressed in the basis.  It is also be reducible to constructing an effective $\Z$-basis for the endomorphism ring of a simple non-ordinary abelian variety. Both problems appear to be hard.

For a simple abelian variety $A$ the center of the division algebra $\End^0 A$ is a CM field $\Q(\pi)$ where $\pi$ is a Weil number (unique up to conjugacy) associated with the Frobenius endomorphism of $A$.  Our analysis shows that an endomorphism $\lambda_D$ determined by a divisor can be effectively expressed as an element in $\Q(\pi)$, if $\lambda_D$ is in the center of $\End^0 A$.  This is an important reason why the discrete logarithm problem can be attacked when $U_1$ is contained in the center of $\End^0 A$.

A consequence of Tate's theorem (\cite{T}, also see Chapter 2 \cite{W}) is that the local invariants of $\End^0 A$ as a division algebra centered at $\Q(\pi)$ is completely determined by $\pi$. The question is, apart from the center, how to explicitly and efficiently determine the structure of $\End^0 A$, or in a smaller scale, the subalgebra generated by two non-commuting elements.   This is an interesting and important problem for further investigation.

\section*{Acknowledgements}
I would like to thank the participants of the AIM workshop on cryptographic multilinear maps for stimulating discussions.  I would also like to thank Dan Boneh and Amit Sahai for helpful discussions during the initial phase of this investigation after the AIM workshop.

\end{document}